\documentclass[12pt]{article}%
\usepackage{amssymb}
\usepackage{amsmath}
\usepackage{amsfonts}
\usepackage{graphicx}%
\begin{document}

\title{{\Large d-Branes in the Stream\ }$^{\text{{\Large \S }}}$}
\author{{\large Thomas Curtright}\\{\large University of Miami}}
\maketitle

\begin{abstract}
Evolution of extended data is considered in various flow problems, using Nambu
brackets as a tool applicable to all cases. \ Extra dimensions, N-brackets,
and extended structures are first employed to linearize the Euler equations.
\ N-bracket induced evolutions of strings, membranes, and other d-branes are
then discussed in detail.

\end{abstract}

\vfill

$^{\text{{\large \S }}}${\footnotesize \ ANL-HEP-CP-03-058 \ \ A talk given at
the International Conference on Nonlinear Evolution Equations and
Applications, Northwestern University, 12 June 2003, based on work with David
Fairlie, University of Durham, and Cosmas Zachos, Argonne National
Laboratory.\ \ Thanks to Ernest Hemingway for the mojito-inspired
title.}\newpage

This talk consists of two more or less independent parts. \ The first part is
a discussion of simple Euler fluid flow, in somewhat unconventional terms.
\ The second part is a consideration of the motion of continuous collections
of points (i.e. branes) in response to a given ambient flow field. \ Although
treated independently, both considerations involve some common mathematical
language: \ the use of Nambu brackets and extended data structures.

\section{Linearizing the Euler-Monge equations}

\paragraph{General results in $n$ dimensions}

Encode the local velocity field for $n$-dimensional flow $\mathbf{u}\left(
\mathbf{x},t\right)  $ into a \emph{non-local structure}, after
\emph{dimension-doubling}. \ Define \cite{CF}%
\begin{gather}
U_{k}\left(  \mathbf{x},t,\mathbf{a}\right)  \equiv\int\cdots\int dq_{1}\cdots
dq_{n}\,\delta\left(  q_{k}-x_{k}\right)  \left(  \frac{e^{a_{k}u_{k}\left(
\mathbf{q},t\right)  }-1}{a_{k}}\right)  \times\nonumber\\
\times\det\left(
\begin{array}
[c]{ccc}%
\frac{\partial}{\partial q_{1}}\left(  \varepsilon\left(  q_{1}-x_{1}\right)
e^{a_{1}u_{1}\left(  \mathbf{q},t\right)  }\right)  & \cdots & \frac{\partial
}{\partial q_{n}}\left(  \varepsilon\left(  q_{1}-x_{1}\right)  e^{a_{1}%
u_{1}\left(  \mathbf{q},t\right)  }\right) \\
\vdots & \ddots & \vdots\\
\frac{\partial}{\partial q_{1}}\left(  \varepsilon\left(  q_{n}-x_{n}\right)
e^{a_{n}u_{n}\left(  \mathbf{q},t\right)  }\right)  & \cdots & \frac{\partial
}{\partial q_{n}}\left(  \varepsilon\left(  q_{n}-x_{n}\right)  e^{a_{n}%
u_{n}\left(  \mathbf{q},t\right)  }\right)
\end{array}
\right)  _{\substack{\text{exclude }k\text{th row}\\\text{and }k\text{th
column}}}\ ,
\end{gather}
where $\;\varepsilon\left(  s\right)  \equiv\pm\frac{1}{2}$ for $s\gtrless
0$\ . \ The local velocity data is given by the extra-dimension boundary
values%
\begin{equation}
u_{k}\left(  \mathbf{x},t\right)  =\lim\limits_{\mathbf{a}\rightarrow
\mathbf{0}}U_{k}\left(  \mathbf{x},t,\mathbf{a}\right)  \ .
\end{equation}
Then with%
\begin{equation}
\mathcal{M}_{n}\left[  \mathbf{u}\right]  \equiv\frac{\partial}{\partial
t}-\sum_{j=1}^{n}u_{j}\frac{\partial}{\partial x_{j}}\;,\;\;\;\;\;\mathcal{H}%
_{n}\equiv\frac{\partial}{\partial t}-\sum_{j=1}^{n}\frac{\partial^{2}%
}{\partial x_{j}\partial a_{j}}\ ,
\end{equation}
we have%
\begin{equation}
\mathcal{M}_{n}\left[  \mathbf{u}\right]  \,u_{i}\left(  \mathbf{x},t\right)
=0
\end{equation}
if and only if \
\begin{equation}
\mathcal{H}_{n}U_{k}\left(  \mathbf{x},t,\mathbf{a}\right)  =0\ .
\end{equation}
The latter equations are solved by the heat kernel method, as given formally
by%
\begin{equation}
U_{k}\left(  \mathbf{x},t,\mathbf{a}\right)  =e^{t\sum_{j=1}^{n}\frac
{\partial^{2}}{\partial x_{j}\partial a_{j}}}\ U_{k}\left(  \mathbf{x}%
,t=0,\mathbf{a}\right)  \ .
\end{equation}
Expansion and evaluation of the RHS yields the formal time power series
solution in terms of the initial data. \ Restricting this series to the extra
dimension boundary, $\mathbf{a}=0,$ we obtain the time power series solution
for $u_{k}\left(  \mathbf{x},t\right)  $ in terms of $u_{k}\left(
\mathbf{x},t=0\right)  $ and its derivatives. \ 

Dimension-doubling here is formally similar to point-splitting in
Schr\"{o}dinger quantum mechanics \cite{CF}. \ But it is nonetheless a bona
fide doubling of dimensions, in the same sense as the Fourier transform of
point-split wave-function bilinears in quantum mechanics yields a Wigner
function on a phase-space whose dimension is twice that of the original
wave-function configuration space. \ 

\paragraph{Nambu brackets}

Classically, N-brackets are defined as Jacobians.%
\begin{equation}
\left\{  A_{1},\cdots,A_{n}\right\}  _{q_{1}\cdots q_{n}}\equiv\frac
{\partial\left(  A_{1},\cdots,A_{n}\right)  }{\partial\left(  q_{1}%
,\cdots,q_{n}\right)  }\ .
\end{equation}
These multi-linear generalizations of the Poisson bracket arise naturally in
the previous construction, since%

\begin{equation}
\delta\left(  q_{k}-x_{k}\right)  \times\det\left(  \cdots\right)
_{\substack{\text{exclude }k\text{th row} \\\text{and }k\text{th column}%
}}=\left\{  \varepsilon\left(  q_{1}-x_{1}\right)  e^{a_{1}u_{1}}%
,\cdots,\varepsilon\left(  q_{k}-x_{k}\right)  ,\cdots,\varepsilon\left(
q_{n}-x_{n}\right)  e^{a_{n}u_{n}}\right\}  _{q_{1}\cdots q_{n}}\ ,
\end{equation}
with the exponential missing from the $k$th entry in the bracket. \ 

In the extra-dimension-boundary limit, the N-bracket becomes a Dirac-delta
equating all components of $\mathbf{x}$ to those of $\mathbf{q}$.%
\begin{equation}
\lim\limits_{\mathbf{a}\rightarrow\mathbf{0}}\left\{  \varepsilon\left(
q_{1}-x_{1}\right)  e^{a_{1}u_{1}},\cdots,\varepsilon\left(  q_{k}%
-x_{k}\right)  ,\cdots,\varepsilon\left(  q_{n}-x_{n}\right)  e^{a_{n}u_{n}%
}\right\}  _{q_{1}\cdots q_{n}}=\delta^{\left(  n\right)  }\left(
\mathbf{q-x}\right)  \ .
\end{equation}
This ensures that $\mathbf{U}$ collapses down to $\mathbf{u}$ on the
$\mathbf{a}\rightarrow\mathbf{0}$ boundary.\bigskip

Note the Nambu bracket is a derivation, and is linear in each argument. \ So
another way to say the above is%
\begin{gather}
\delta\left(  q_{k}-x_{k}\right)  e^{a_{k}u_{k}}\det\left(  \cdots\right)
_{\substack{\text{exclude }k\text{th row} \\\text{and }k\text{th column}%
}}=\left\{  \varepsilon\left(  q_{1}-x_{1}\right)  e^{a_{1}u_{1}}%
,\cdots,\varepsilon\left(  q_{n}-x_{n}\right)  e^{a_{n}u_{n}}\right\}
_{q_{1}\cdots q_{n}}\nonumber\\
-a_{k}\,\varepsilon\left(  q_{k}-x_{k}\right)  \,e^{a_{k}u_{k}}\left\{
\varepsilon\left(  q_{1}-x_{1}\right)  e^{a_{1}u_{1}},\cdots,u_{k}%
,\cdots,\varepsilon\left(  q_{n}-x_{n}\right)  e^{a_{n}u_{n}}\right\}
_{q_{1}\cdots q_{n}}\ .
\end{gather}

\paragraph{A trivial example in one-dimension}

As the simplest, local example, consider%
\begin{equation}
U\left(  x,t,a\right)  \equiv\frac{e^{au\left(  x,t\right)  }-1}%
{a}\ ,\ \ \ u\left(  x,t\right)  =\frac{1}{a}\ln\left(  1+aU\left(
x,t,a\right)  \right)  =\lim\limits_{a\rightarrow0}U\left(  x,t,a\right)  \ .
\end{equation}
Then%
\begin{equation}
\frac{\partial}{\partial t}U\left(  x,t,a\right)  =\frac{\partial^{2}%
}{\partial a\partial x}U\left(  x,t,a\right)
\end{equation}
if and only if%
\begin{equation}
\frac{\partial}{\partial t}u\left(  x,t\right)  =u\left(  x,t\right)
\frac{\partial}{\partial x}u\left(  x,t\right)  \ .
\end{equation}
Moreover,%
\begin{equation}
U\left(  x,t,a\right)  =e^{t\,\frac{\partial^{2}}{\partial a\partial x}%
}\ U\left(  x,t=0,a\right)
\end{equation}
immediately gives the time power series solution from initial data%
\begin{equation}
u\left(  x,t\right)  =\lim\limits_{a\rightarrow0}e^{t\,\frac{\partial^{2}%
}{\partial a\partial x}}\ \left(  \frac{e^{au\left(  x\right)  }-1}{a}\right)
=\sum\limits_{j=0}^{\infty}\frac{t^{j}}{\left(  1+j\right)  !}\frac{d^{j}%
}{dx^{j}}\left(  u\left(  x\right)  \right)  ^{1+j}\ .
\end{equation}
So, for example, an initial linear profile, $u\left(  x\right)  =\alpha+\beta
x$, evolves into $u\left(  x,t\right)  =\left(  \alpha+\beta x\right)
/\left(  1-\beta t\right)  $. \ For a given interval in $u$, this steepens as
time increases, for $\beta>0$, and becomes vertical at $t_{\text{break}%
}=1/\beta$. \ Other simple breaking wave examples can be found in \cite{CF}
(or at http://curtright.com/waves.html).

\paragraph{Two-dimensions and some non-locality}

Define the extended data structures ($U_{1}\equiv U,\ U_{2}\equiv V$, etc.)%
\begin{align}
U\left(  x,y,t,a,b\right)   &  \equiv\int dr\;\varepsilon\left(  y-r\right)
\;e^{au\left(  x,r,t\right)  +bv\left(  x,r,t\right)  }\;\frac{\partial
u\left(  x,r,t\right)  }{\partial r}\ ,\nonumber\\
V\left(  x,y,t,a,b\right)   &  \equiv\int dq\;\varepsilon\left(  x-q\right)
\;e^{au\left(  q,y,t\right)  +bv\left(  q,y,t\right)  }\;\frac{\partial
v\left(  q,y,t\right)  }{\partial q}\ .
\end{align}
Then%
\begin{equation}
\mathcal{H}_{2}U=0=\mathcal{H}_{2}V \label{2Heat}%
\end{equation}
if and only if \
\begin{equation}
\mathcal{M}_{2}u=0=\mathcal{M}_{2}v\ . \label{2EM}%
\end{equation}
Again, note the extra-dimension boundary limits%
\begin{equation}
u\left(  x,y,t\right)  =\lim\limits_{a,b\rightarrow0}U\left(
x,y,t,a,b\right)  \;,\;\;\;v\left(  x,y,t\right)  =\lim\limits_{a,b\rightarrow
0}V\left(  x,y,t,a,b\right)  \ .
\end{equation}
The demonstration of the equivalence of (\ref{2Heat}) and (\ref{2EM}) goes as
follows. \ Direct calculation gives%
\begin{multline}
\mathcal{H}_{2}U\left(  x,y,t,a,b\right)  =e^{au\left(  x,y,t\right)
+bv\left(  x,y,t\right)  }\mathcal{M}_{2}u\left(  x,y,t\right) \nonumber\\
+b\int dr\;\varepsilon\left(  y-r\right)  \;e^{au\left(  x,r,t\right)
+bv\left(  x,r,t\right)  }\left(  \frac{\partial u\left(  x,r,t\right)
}{\partial r}\mathcal{M}_{2}v\left(  x,r,t\right)  -\frac{\partial v\left(
x,r,t\right)  }{\partial r}\mathcal{M}_{2}u\left(  x,r,t\right)  \right)  \ .
\end{multline}
Similarly for $V$. \ So then $\mathcal{M}_{2}u=0=\mathcal{M}_{2}v$ clearly
implies $\mathcal{H}_{2}U=0$. \ The converse follows by using the obvious
limit
\begin{equation}
\lim\limits_{a,b\rightarrow0}\mathcal{H}_{2}U\left(  x,y,t,a,b\right)
=\mathcal{M}_{2}u\left(  x,y,t\right)  \ ,
\end{equation}
along with the similar relation for $V$. \ {\scriptsize QED}

\paragraph{Three-dimensions and Poisson brackets}

($U_{1}\equiv U,\ U_{2}\equiv V,\ U_{3}=W$, etc.)%
\begin{equation}
\mathcal{H}_{3}U=\mathcal{H}_{3}V=\mathcal{H}_{3}W=0 \label{3Heat}%
\end{equation}
if and only if \
\begin{equation}
\mathcal{M}_{3}u=\mathcal{M}_{3}v=\mathcal{M}_{3}w=0\ , \label{3EM}%
\end{equation}
where%
\begin{align}
U\left(  x,y,z,t,a,b,c\right)   &  \equiv\int dr\,\varepsilon\left(
y-r\right)  \,e^{au+bv+cw}\;\frac{\partial u\left(  x,r,z,t\right)  }{\partial
r}\nonumber\\
&  -c\iint drds\,\varepsilon\left(  y-r\right)  \,\varepsilon\left(
z-s\right)  \,e^{au+bv+cw}\left\{  u,w\right\}  _{rs}\left(  x,r,s,t\right)
\ ,
\end{align}%
\begin{align}
V\left(  x,y,z,t,a,b,c\right)   &  \equiv\int ds\,\varepsilon\left(
z-s\right)  \,e^{au+bv+cw}\;\frac{\partial v\left(  x,y,s,t\right)  }{\partial
s}\nonumber\\
&  -a\iint dqds\,\varepsilon\left(  x-q\right)  \,\varepsilon\left(
z-s\right)  \,e^{au+bv+cw}\left\{  v,u\right\}  _{sq}\left(  q,y,s,t\right)
\ ,
\end{align}%
\begin{align}
W\left(  x,y,z,t,a,b,c\right)   &  \equiv\int dq\,\varepsilon\left(
x-q\right)  \,e^{au+bv+cw}\;\frac{\partial w\left(  q,y,z,t\right)  }{\partial
q}\nonumber\\
&  -b\iint dqdr\,\varepsilon\left(  x-q\right)  \,\varepsilon\left(
y-r\right)  \,e^{au+bv+cw}\left\{  w,v\right\}  _{qr}\left(  q,r,z,t\right)
\ ,
\end{align}
and where the Poisson bracket is given as usual by%
\begin{equation}
\left\{  u,v\right\}  _{rs}=\frac{\partial u}{\partial r}\frac{\partial
v}{\partial s}-\frac{\partial u}{\partial s}\frac{\partial v}{\partial r}\;.
\end{equation}
Once again the proof\ of the equivalence of (\ref{3Heat}) and (\ref{3EM}) is
given by direct calculation. \ Higher dimensions lead to higher rank Nambu
brackets, as expressed by the general $n$ result quoted above.

\section{Nambu evolution of d-branes}

Consider flow-driven motion in a local velocity field $\mathbf{u}\left[
x\right]  .$ For points moving along with the flow\footnote{There is one
caveat to keep in mind here. \ It is tempting to think of $\tau$ as
conventional time $t$, defined universally. \ But this need not be so. \ In
fact, in some situations it can be misleading to think of $\tau$ this way.
\ For streamline fluid flow, e.g., $\tau$ could be allowed to vary from one
streamline to the next such that evolution along each streamline is governed
by that streamline's own individual clock. \ In this sense, $\tau$ is more
like proper time in relativity. \ In the context of Nambu mechanics in
phase-space, $\tau$ is generally related to $t$ by a \emph{dynamical scale
factor}, and therefore \emph{cannot} be simply identified with universal time
across all dynamical sectors \cite{CZ}. \ That is to say, $\tau$ may be
trajectory dependent. \ For this and other reasons to become apparent, we will
call $\tau$ \textquotedblleft Nambu time.\textquotedblright}%
\begin{equation}
dx^{i}=u^{i}\left[  x\right]  d\tau\;. \label{GoWithTheFlow}%
\end{equation}
This deceptively simple description is general enough to incorporate the
phase-space evolution of any Hamiltonian dynamical system, upon identifying
pairs of variables as canonically conjugate, say $\left(  x^{j},x^{j+1}%
\right)  \rightarrow\left(  x^{j},p^{j}\right)  $. \ 

Is there an elegant way to describe the motion of an extended, continuous
collection of points moving with the flow? \ For example, consider (vortex)
strings, (surface interface) membranes, or higher dimensional manifolds of
points immersed in fluids flowing in higher dimensions. \ The latter higher
dimensional generalization is most important for dynamics in phase-space,
which for a system with $N$ degrees of freedom is $2N$ dimensional. \ That is
to say, if we parameterize the extended set of points for a given $\tau$ as%
\begin{equation}
x\left(  \tau,\alpha_{1},\alpha_{2},\cdots,\alpha_{d}\right)  \ ,
\end{equation}
then we have a d-dimensional \emph{data-brane},\ or \textquotedblleft
d-brane\textquotedblright\ for short. \ What is the most economical way to
describe the collective evolution of this extended set of data?

A suggested answer to both these rhetorical questions is as follows. \ If the
ambient velocity field is given by a Nambu bracket, as%
\begin{equation}
u^{i}\left[  x\right]  =\left\{  x^{i},I_{1},\cdots,I_{n-1}\right\}  \;,
\label{NambuField}%
\end{equation}
where the $I$s are \textquotedblleft flow invariants,\textquotedblright\ then
we have satisfied a sufficient condition to obtain an \emph{exact} $n$-form,
$\Omega=d\Lambda$, that provides a minimal coupling of the d-brane to the
velocity field, for the case $d=n-2$. \ The individual points are governed by
Nambu dynamics \cite{Nambu} in this case, with%
\begin{equation}
\frac{dx^{i}}{d\tau}=\left\{  x^{i},I_{1},\cdots,I_{n-1}\right\}  \;.
\end{equation}
Examples of such dynamical systems in phase-space are abundantly provided by
\emph{superintegrable systems}. \ An $N$ degree-of-freedom system, with
phase-space of dimension $2N$,\ is \emph{maximally superintegrable} and
described as above when $n=2N$ with a corresponding set of $2N-1$ invariants,
$I_{i}$. \ In this $2N$ phase-space, there can be no more than $2N-1$
invariants. \ These ideas are extensively discussed in \cite{CZ}.

As already noted, the preferred continuous collection of points, i.e. the
minimally coupled d-brane, is of dimension $d=n-2$ when the velocity field is
given by the above Nambu bracket. \ The explicit $n$-form that provides the
minimal coupling of the brane to the velocity field is then%
\begin{align}
\Omega &  =\varepsilon_{i_{1}\cdots i_{n}}\left(  dx^{i_{1}}-u^{i_{1}}\left[
x\right]  d\tau\right)  \wedge\cdots\wedge\left(  dx^{i_{n}}-u^{i_{n}}\left[
x\right]  d\tau\right) \nonumber\\
&  =\varepsilon_{i_{1}\cdots i_{n}}dx^{i_{1}}\wedge\cdots\wedge dx^{i_{n}%
}-n\varepsilon_{i_{1}\cdots i_{n}}u^{i_{1}}\left[  x\right]  d\tau\wedge
dx^{i_{2}}\cdots\wedge dx^{i_{n}}\nonumber\\
&  =n!\left(  dx^{1}\wedge\cdots\wedge dx^{n}-d\tau\wedge dI_{1}\wedge
\cdots\wedge dI_{n-1}\right)  \;. \label{GoodForm}%
\end{align}
The transition from the first to the second line here is trivial, while the
detailed steps leading from the second to the third lines are given below, in
(\ref{TheDetails}). \ Even without detailed index manipulations, however, the
form of the third line is transparent given (\ref{NambuField}) and the form of
the second line, and vice versa. \ 

For instance, to go from the second to the third lines, first note that
$\Omega$ is first-order in $u$. \ Then, since $u^{i}$ is multi-linear and
totally antisymmetric in the first partial $x^{j\neq i}$ derivatives of the
$I$s, the terms linear in $u^{i}$ must sum to give $dI_{1}\wedge\cdots\wedge
dI_{n-1}$. \ The only thing that is perhaps not obvious is the relative
coefficient between the two exact $n$-forms that appear in the third line of
(\ref{GoodForm}). \ This relative coefficient is carefully and directly
determined below, in (\ref{TheDetails}). \ An alternate albeit indirect
verification of the relative coefficient, as written in (\ref{GoodForm}), was
first given in \cite{Tahktajan} through Tahktajan's demonstration that the
obvious extrema of the integrals of the first and second lines of
(\ref{GoodForm}) are also extrema of integrals of the third line. \ This test
would fail for other relative coefficients.

Further consideration of such integrals is warranted. \ We want $\Omega$ to be
exact\footnote{A necessary condition for $\Omega$ to be exact is that it be
closed. \ This implies the flow should be divergenceless, $\nabla\cdot
$\textbf{$u$}$=0,$ as $\varepsilon_{i_{1}\cdots i_{n}}du^{i_{1}}\left[
x\right]  \wedge dx^{i_{2}}\cdots\wedge dx^{i_{n}}=\varepsilon_{i_{1}\cdots
i_{n}}\partial_{j}u^{i_{1}}dx^{j}\wedge dx^{i_{2}}\cdots\wedge dx^{i_{n}%
}=\frac{1}{n}\left(  \partial_{j}u^{j}\right)  \varepsilon_{i_{1}\cdots i_{n}%
}dx^{i_{1}}\wedge dx^{i_{2}}\cdots\wedge dx^{i_{n}}\;.$} to trivially reduce
$\int_{M_{n}}\Omega$ to $\int_{\partial M_{n}}\Lambda$, i.e. to a
\textquotedblleft boundary action,\textquotedblright\ that describes evolution
of the d-brane. \ That action will be manifestly extremal on the solution set
defined by (\ref{GoWithTheFlow}), given the multilinear form of $\Omega$ in
terms of those first order equations of motion. \ The action that governs the
evolution of the d-brane (for $n=3$, see \cite{Estabrook}, and\ for general
$n$, see \cite{Tahktajan}, especially Remark 9) is%
\begin{equation}
A=\int_{M_{n}}\Omega=n!\int_{\partial M_{n}}\left(  x^{1}\wedge dx^{2}%
\wedge\cdots\wedge dx^{n}+I_{1}d\tau\wedge dI_{2}\wedge\cdots\wedge
dI_{n-1}\right)  \;.
\end{equation}
The latter form of the action is an integral over the $n-1$ dimensional
world-volume swept out by the evolving d-brane, confirming that at any instant
the d-brane is of dimension $n-2$. \ Parameterizing the world-volume by
coordinates $\alpha_{1},\cdots,\alpha_{n-2}$, as well as $\tau$, as already
chosen, the action is%
\begin{equation}
A=\int_{\partial M_{n}}d\tau d\alpha_{1}\cdots d\alpha_{n-2}\left[
\begin{array}
[c]{c}%
\left(  n-1\right)  !\;\varepsilon_{i_{1}\cdots i_{n}}x^{i_{1}}\partial_{\tau
}x^{i_{2}}\partial_{\alpha_{1}}x^{i_{3}}\cdots\partial_{\alpha_{n-2}}x^{i_{n}%
}\\
+\left(  n-2\right)  !\ n\;\varepsilon^{j_{1}\cdots j_{n-1}}I_{j_{1}}%
\partial_{\alpha_{1}}I_{j_{2}}\cdots\partial_{\alpha_{n-2}}I_{j_{n-1}}%
\end{array}
\right]  \ .
\end{equation}
NB \ In this last expression, all $i$s are summed from $1$ to $n$, but all
$j$s are summed from $1$ to $n-1$.

The equations of motion that follow from extremizing this action are then
projected versions of the original (\ref{GoWithTheFlow}), namely%
\begin{equation}
\varepsilon_{i_{1}\cdots i_{n}}\left(  \partial_{\tau}x^{i_{2}}-u^{i_{2}%
}\right)  \partial_{\alpha_{1}}x^{i_{3}}\cdots\partial_{\alpha_{n-2}}x^{i_{n}%
}=0\;. \label{FlowWithTheGo}%
\end{equation}
This projected form of the equations of motion (\ref{GoWithTheFlow}) allows
for \emph{arbitrary} reparameterizations of the $\left(  n-2\right)  $-brane
surface, at any given $\tau$, including $\tau$-dependent choices for the
$\alpha_{j}$. \ Through this reparameterization freedom, as emphasized by
Regge et al. \cite{Regge}, and subsequently by Tahktajan \cite{Tahktajan}
(especially Remark 8), there exists a $\tau$-dependent parameterization such
that (\ref{GoWithTheFlow}) is fully recovered. \ However, in more
geometrical/physical terms, the evolution of a point according to
(\ref{GoWithTheFlow}) is fully recovered just by considering the
(\ref{FlowWithTheGo})-driven evolution of $n-1$ intersecting branes,
appropriately configured to intersect at the point in question. \ This is most
easily visualized for the string case, with $n=3$. \ A point defined by the
intersection of two appropriately chosen strings, with each string moving
according to (\ref{FlowWithTheGo}), will itself evolve according to
(\ref{GoWithTheFlow}).

What is most interesting\ in all this is the incorporation of interactions
through the use of Nambu brackets, particularly for the case of even
dimensional phase-space d-branes for superintegrable systems. \ This elegant
geometrical characterization of integrable systems with a maximal number of
invariants has \emph{not} been fully appreciated previously, I believe.
\ Also, this novel geometrical view holds considerable promise for
understanding the quantization of such maximally superintegrable systems --
promise made all the more possible by recent work on the quantization of the
Nambu bracket using both traditional Hilbert space operator and non-Abelian
deformation methods of quantization \cite{CZ}.

Specification of the dynamics through the $n$-form $\Omega$ does not by itself
endow the brane with any inherent stability or cohesion. \ That is, the brane
is \emph{tensionless}\ so far as the above action dictates. \ The data
comprising the brane evolves smoothly and remains contiguous without tearing
only through the courtesy of the driving velocity field. \ If the latter is
smooth enough, then so is the evolution of the extended data set. \ It may
move forward along with the ambient flow and give the appearance of stability,
much as a thin sheet of smoke, dust, or snow blown by the wind may stay
together if the driving flow of air allows it. \ Moreover, for a $\tau
$-independent but otherwise arbitrarily chosen parameterization, the d-brane
responds similarly to an untethered, tensionless, \textquotedblleft
teflon-coated\textquotedblright\ sail (for the $n=4$ case, with more
challenging mental pictures for generalizations to higher $n$). \ Only the
flow component perpendicular to the surface has any apparent effect on it.
\ This would suggest that any shear\ in the ambient flow will cause all
impacted sections of the surface to rotate and become parallel to $\mathbf{u}
$ (i.e. tangential). \ Once parallel to $\mathbf{u}$, the surface sections
will cease being moved extrinsically by the velocity field, although intrinsic
$\tau$-dependent reparameterizations can be used to maintain the tangential
motion inherent in (\ref{GoWithTheFlow}). \ At least, that is the case for
motion due to the action of the $n$-form alone.

For classification purposes, it is sufficient to consider just the free
kinetic portion of the $n$-form (Hopf term).%
\begin{equation}
\omega=\varepsilon_{i_{1}\cdots i_{n}}dx^{i_{1}}\wedge\cdots\wedge dx^{i_{n}%
}=n!\left(  dx^{1}\wedge\cdots\wedge dx^{n}\right)  \;.
\end{equation}
This alone is exact, so we obtain only a boundary action from its integral.%
\begin{equation}
\int_{M_{n}}\omega=\int_{\partial M_{n}}\varepsilon_{i_{1}\cdots i_{n}%
}x^{i_{1}}dx^{i_{2}}\wedge\cdots\wedge dx^{i_{n}}=\left(  n-1\right)  !\int
d\tau d\alpha_{1}\cdots d\alpha_{n-2}\,\varepsilon_{i_{1}\cdots i_{n}}%
x^{i_{1}}\partial_{\tau}x^{i_{2}}\partial_{\alpha_{1}}x^{i_{3}}\cdots
\partial_{\alpha_{n-2}}x^{i_{n}}\;,
\end{equation}
where in the last step we have again used the parameterization of the $\left(
n-1\right)  $-dimensional world-volume swept out by the evolving $\left(
n-2\right)  $-brane. \ We have singled out one of the parameters as the Nambu
time, $\tau$. \ So the free term may be used to classify the various extended
structures, as for example:
\begin{align}
\int_{M_{2}}\omega &  =\int_{\partial M_{2}}\varepsilon_{ij}x^{i}dx^{j}=\int
d\tau\,\varepsilon_{ij}x^{i}\partial_{\tau}x^{j}\text{ \ \ \ \ point
particle,}\\
\int_{M_{3}}\omega &  =\int_{\partial M_{3}}\varepsilon_{ijk}x^{i}dx^{j}\wedge
dx^{k}=2\int d\tau d\alpha\,\varepsilon_{ijk}x^{i}\partial_{\tau}x^{j}%
\partial_{\alpha}x^{k}\text{ \ \ \ \ string,}\nonumber\\
\int_{M_{4}}\omega &  =\int_{\partial M_{4}}\varepsilon_{ijkl}x^{i}%
dx^{j}\wedge dx^{k}\wedge dx^{l}=6\int d\tau d\alpha d\beta\,\varepsilon
_{ijkl}x^{i}\partial_{\tau}x^{j}\partial_{\alpha}x^{k}\partial_{\beta}%
x^{l}\text{ \ \ \ \ membrane,}\nonumber\\
\int_{M_{2N}}\omega &  =\left(  2N\right)  !\int_{M_{2N}}dx^{1}\wedge
dp^{1}\wedge dx^{2}\wedge dp^{2}\wedge\cdots\wedge dx^{N}\wedge dp^{N}%
=2^{N}\left(  2N-1\right)  !!\int_{M_{2N}}\left(  \omega_{2}\right)
^{N}\nonumber\\
&  =\left(  2N\right)  !\int_{M_{2N-1}=\partial M_{2N}}x^{1}\wedge
dp^{1}\wedge\cdots\wedge dx^{N}\wedge dp^{N}\text{ \ \ \ \ maximal phase-space
d-brane,}\nonumber
\end{align}
where in the next to last line, $\omega_{2}=\sum_{j}dx^{j}\wedge dp^{j}$ is
the canonical two-form on the phase-space.

Take the simplest case of a particle in phase-space undergoing Hamiltonian
flow, with $n=2$. \ The phase-space velocity field in this case is given by
\begin{equation}
u^{x}=\frac{\partial}{\partial p}H\;,\;\;\;u^{p}=-\frac{\partial}{\partial
x}H\;,\;\;\;\frac{\partial}{\partial x}u^{x}+\frac{\partial}{\partial p}%
u^{p}=0\;.
\end{equation}
The additional term is then%
\begin{equation}
\varepsilon_{i_{1}\cdots i_{n}}u^{i_{1}}\left[  x\right]  d\tau\wedge
dx^{i_{2}}\cdots\wedge dx^{i_{n}}\;_{\overrightarrow{n=2}}\;d\tau\wedge\left(
u^{x}dp-u^{p}dx\right)  =d\tau\wedge\left(  \partial_{p}Hdp+\partial
_{x}Hdx\right)  =d\tau\wedge dH\;,
\end{equation}
the expected exact 2-form, well-known from Hamiltonian dynamics. \ So the
action is the usual%
\begin{equation}
A=\int_{M_{2}}\Omega=\int_{\partial M_{2}}xdp-pdx+2Hd\tau=2\int_{\partial
M_{2}}Hd\tau-pdx\;.
\end{equation}
The method always works for the point-particle case when the flow is Hamiltonian.

Take the second simplest case of a \textquotedblleft vortex
string\textquotedblright\ (see Regge et al. \cite{Regge}), with $n=3$.%
\begin{equation}
\varepsilon_{i_{1}\cdots i_{n}}u^{i_{1}}\left[  x\right]  d\tau\wedge
dx^{i_{2}}\cdots\wedge dx^{i_{n}}\;_{\overrightarrow{n=3}}\;2\ d\tau
\wedge\left(  u^{x}dy\wedge dz+u^{y}dz\wedge dx+u^{z}dx\wedge dy\right)  \;.
\end{equation}
When the velocity field is given by a Nambu bracket, with \emph{invariant
generating entries} that we shall designate $H$ and $L$, we obtain:%
\begin{align}
u^{x}  &  =\left\{  x,H,L\right\}  =\partial_{y}H\partial_{z}L-\partial
_{z}H\partial_{y}L\;,\nonumber\\
u^{y}  &  =\left\{  y,H,L\right\}  =\partial_{z}H\partial_{x}L-\partial
_{x}H\partial_{z}L\;,\nonumber\\
u^{z}  &  =\left\{  z,H,L\right\}  =\partial_{x}H\partial_{y}L-\partial
_{y}H\partial_{x}L\;.
\end{align}
For the velocity field around a vortex, these two invariants may in fact be
written in terms of the Clebsch potentials \cite{NambuClebsch,Rylov}. \ In any
case,%
\begin{gather}
u^{x}dy\wedge dz+u^{y}dz\wedge dx+u^{z}dx\wedge dy\nonumber\\
=\left(  \partial_{y}H\partial_{z}L-\partial_{z}H\partial_{y}L\right)
dy\wedge dz+\left(  \partial_{z}H\partial_{x}L-\partial_{x}H\partial
_{z}L\right)  dz\wedge dx+\left(  \partial_{x}H\partial_{y}L-\partial
_{y}H\partial_{x}L\right)  dx\wedge dy\nonumber\\
=\left(  \partial_{x}Hdx+\partial_{y}Hdy+\partial_{z}Hdz\right)  \wedge\left(
\partial_{x}Ldx+\partial_{y}Ldy+\partial_{z}Ldz\right) \nonumber\\
=dH\wedge dL\;.
\end{gather}
Thus the interaction with the velocity field becomes%
\begin{equation}
\varepsilon_{i_{1}\cdots i_{n}}u^{i_{1}}\left[  x\right]  d\tau\wedge
dx^{i_{2}}\cdots\wedge dx^{i_{n}}\;_{\overrightarrow{n=3}}\;2\ d\tau\wedge
dH\wedge dL\;,
\end{equation}%
\begin{equation}
\int_{M_{3}}\Omega=6\int_{M_{3}}dx\wedge dy\wedge dz-d\tau\wedge dH\wedge
dL=6\int_{\partial M_{3}}xdy\wedge dz-HdL\wedge d\tau\;.
\end{equation}

Next, consider the details of the general case. \ We will use the generalized
Kronecker symbol and its trace (both $i$s and $j$s are summed from $1$ to $n$
here).%
\begin{align}
\varepsilon_{i_{1}\cdots i_{n}}\varepsilon_{j_{1}\cdots j_{n}}  &
=\delta_{j_{1}\cdots j_{n}}^{_{i_{1}\cdots i_{n}}}=\delta_{j_{1}}^{_{i_{1}}%
}\times\delta_{j_{2}j_{3}\cdots j_{n}}^{_{i_{2}i_{3}\cdots i_{n}}}%
-\delta_{j_{2}}^{_{i_{1}}}\times\delta_{j_{1}j_{3}\cdots j_{n}}^{_{i_{2}%
i_{3}\cdots i_{n}}}+-\text{ \ \ \ \ (}n\text{ terms),}\nonumber\\
\delta_{j_{2}\cdots j_{n}}^{_{i_{2}\cdots i_{n}}}  &  =\delta_{i_{1}%
j_{2}\cdots j_{n}}^{_{i_{1}i_{2}\cdots i_{n}}}=\delta_{j_{1}}^{i_{1}}%
\times\delta_{i_{1}\cdots i_{n}}^{j_{1}\cdots j_{n}}\;.
\end{align}
Note from the examples the point is always to make an exact form out of%
\begin{equation}
\varepsilon_{i_{1}\cdots i_{n}}u^{i_{1}}\left[  x\right]  dx^{i_{2}}%
\cdots\wedge dx^{i_{n}}\;.
\end{equation}
But when the velocity field is given by an $n$-bracket, $u^{i_{1}}=\left\{
x^{i_{1}},I_{1},\cdots,I_{n-1}\right\}  $ this becomes%
\begin{align}
\varepsilon_{i_{1}\cdots i_{n}}\left\{  x^{i_{1}},I_{1},\cdots,I_{n-1}%
\right\}  \,dx^{i_{2}}\wedge\cdots\wedge dx^{i_{n}}  &  =\varepsilon
_{i_{1}\cdots i_{n}}\varepsilon_{j_{1}\cdots j_{n}}\,\partial_{j_{1}}x^{i_{1}%
}\partial_{j_{2}}I_{1}\cdots\partial_{j_{n}}I_{n-1}dx^{i_{2}}\wedge
\cdots\wedge dx^{i_{n}}\nonumber\\
&  =\varepsilon_{i_{1}\cdots i_{n}}\varepsilon_{j_{1}\cdots j_{n}}%
\delta_{j_{1}}^{i_{1}}\,\partial_{j_{2}}I_{1}\cdots\partial_{j_{n}}%
I_{n-1}dx^{i_{2}}\wedge\cdots\wedge dx^{i_{n}}\nonumber\\
&  =\delta_{i_{1}\cdots i_{n}}^{j_{1}\cdots j_{n}}\delta_{j_{1}}^{i_{1}%
}\,\partial_{j_{2}}I_{1}\cdots\partial_{j_{n}}I_{n-1}dx^{i_{2}}\wedge
\cdots\wedge dx^{i_{n}}\nonumber\\
&  =\delta_{j_{2}\cdots j_{n}}^{_{i_{2}\cdots i_{n}}}\,\partial_{j_{2}}%
I_{1}\cdots\partial_{j_{n}}I_{n-1}dx^{i_{2}}\wedge\cdots\wedge dx^{i_{n}%
}\nonumber\\
&  =\left(  n-1\right)  !\,\partial_{i_{2}}I_{1}dx^{i_{2}}\wedge\cdots
\wedge\partial_{i_{n}}I_{n-1}dx^{i_{n}}\nonumber\\
&  =\left(  n-1\right)  !\,dI_{1}\wedge\cdots\wedge dI_{n-1}\;.
\label{TheDetails}%
\end{align}
That's all there is to it. \ In summary, there are two steps needed:%
\begin{equation}
\varepsilon_{i_{1}\cdots i_{n}}u^{i_{1}}\left[  x\right]  dx^{i_{2}}%
\cdots\wedge dx^{i_{n}}=\varepsilon_{i_{1}\cdots i_{n}}\left\{  x^{i_{1}%
},I_{1},\cdots,I_{n-1}\right\}  dx^{i_{2}}\wedge\cdots\wedge dx^{i_{n}%
}=\left(  n-1\right)  !dI_{1}\wedge\cdots\wedge dI_{n-1}\;.
\end{equation}
Thus, for such velocity fields given by Nambu brackets, we recover
(\ref{GoodForm}).

A more complete recount of the history of the ambient \textquotedblleft
flowing fluid\textquotedblright\ picture of the classical dynamics of extended
objects would take note of the following. \ Following the seminal paper of
Nambu, the classical theory in the general case was worked out in various
stages, by Estabrook, by Lund, Rasetti, and Regge, by Tahktajan, and by Matsuo
and Shibusa. \ While extended objects were not explicitly discussed, this
picture was nevertheless initiated, formally, as a generalization of the usual
Hamiltonian 2-form framework to higher-dimensional forms by Estabrook
\cite{Estabrook} in the context of Nambu mechanics \cite{Nambu}. \ A physical
interpretation as an extended object moving in a three-dimensional fluid was
given by Regge et al. \cite{Regge}, for the case of vortex strings, but
without using N-brackets. \ The whole business was mathematically codified for
general $n$ by Tahktajan \cite{Tahktajan}, using N-brackets, but without any
additional physical interpretation, and with some unnecessary complication
involving an \emph{extended} phase-space picture. \ More recently, Matsuo and
Shibusa \cite{Matsuo} have developed further the ideas of Regge et al. to
higher $n$, and Pioline \cite{Pioline} has given a summary of past
developments, with emphasis on strings and membranes, while stressing the case
of 3-forms. \ Even more recently, Curtright and Zachos \cite{CZ} have
discussed at length the Nambu formalism in \emph{conventional} phase-space
contexts for superintegrable systems, have emphasized the differences between
even and odd N-brackets, have pointed out the particularly nice properties of
implementing Nambu mechanics with 4-brackets for models with $su\left(
2\right)  $ invariance, and have carefully discussed the quantization of the
formalism using conventional Hilbert space and/or deformation quantization
methods. \ Finally, in a conference talk reviewing \cite{CZ}, Zachos has
discussed 2-branes using 4-brackets \cite{Zachos}.

Only classical considerations were made above. \ However, the resulting mental
picture evoked by the formalism is somewhere intermediate between the
classical Hamiltonian evolution of ideal points, and the quantum evolution of
globally well-behaved distributions (Wigner functions, say, defined throughout
the full phase space) with support of codimension zero. \ Here we are led to
envision distributions of data on surfaces of non-zero codimension, evolved
quite elegantly through the use of a single Nambu bracket in such a way that
the implicit evolution of ideal points comprising the surface is not
necessarily just uniform time evolution, but rather evolution involving a
range of time scales dictated by the breadth of the extended distribution in
the various invariants. \ 

That is to say, the complete extended object evolves smoothly and uniformly
under the passage of a single Nambu time, $\tau$. \ But each individual point
comprising the object may evolve according to its own time scale such that a
spectrum of conventional individual particle times $t$ pass during the
evolution. \ Aside from being at least implicit in the references cited
earlier, the only similar classical precursor of such collectively evolved
data, as far as we know, arises in the formalism of the phase-space
surface-of-section used in the exploration of chaos and to test for
integrability. \ In that formalism, the points making-up the initial surface
are evolved individually, for example by Hamiltonian methods (see, for
example, \cite{Gutzwiller}). \ When following individual trajectories of
points initially on the surface of section, the times required to return to
the initial surface may vary between the various trajectories. \ Under Nambu
time evolution, the analogue of the surface of section would evolve as a
coherent whole for each Nambu time increment. \ 

The quantization of the d-brane dynamics described here is under
investigation. \ Early results on quantization issues for 1-brane vortices are
described in \cite{Regge}. \ A follow-up infusion of general, more rigorous
results for quantized 2-branes of vorticity is given in \cite{Goldin}. \ Still
more recent results for 1-branes are in \cite{Fischer,Spelio}. \ None of these
other studies of the quantization of vorticity and 1-branes makes particular
use of quantum Nambu brackets, however. \ Such usage seems far less
problematic today than it did, say, one year ago due to the developments in
\cite{CZ}. \ When N-brackets are quantized, they have many surprising
features, not the least of which is that the time-scale becomes dynamical in
all but one known special case, when applied to the evolution of
point-particle Wigner functions. \ This is one feature that naturally
associates quantum N-brackets to the evolution of d-branes described above,
and strongly suggests that the latter can provide a convenient bridge between
classical and quantum behavior. \ A full report on this subject will have to
be given elsewhere.

\subparagraph{Acknowledgements:}

For their collaborations and/or insightful discussions, I thank David Fairlie,
Yoichiro Nambu, and Cosmas Zachos, the latter especially for his retrenchant
remarks and for furthering my appreciation of reference \cite{Estabrook}.
\ This research was supported in part by NSF Award 0073390, and in part by the
US Department of Energy, Division of High Energy Physics, under contract
W-31-109-ENG-38, while the author was visiting Argonne National Laboratory,
Argonne, IL 60439. \ I thank the Argonne Particle Theory Group for its warm
hospitality.

\end{document}